\documentclass[12pt]{article}

\usepackage[english]{babel}
\usepackage[utf8]{inputenc}
\usepackage[T1]{fontenc}
\usepackage{amsmath}
\usepackage{cite}
\usepackage{url}
\usepackage{authblk}
\usepackage[bookmarks,colorlinks=false]{hyperref}

\newcounter{mnotecount}[section]

\DeclareMathOperator*{\wlim}{w-lim}

\begin{document}
\title{Inhomogeneity effect in Wainwright-Marshman space-times}
\author[1]{Sebastian J. Szybka}
\author[1,3]{Krzysztof Głód}
\author[2]{Michał J. Wyrębowski}
\author[1]{Alicja Konieczny}
\affil[1]{Astronomical Observatory, Jagellonian University}
\affil[2]{Institute of Physics, Jagellonian University}
\affil[3]{Copernicus Center for Interdisciplinary Studies}
\date{}
\maketitle{}
\begin{abstract}
Green and Wald have presented a mathematically rigorous framework to study, within general relativity, the effect of small-scale inhomogeneities on the global structure of space-time. 
The framework relies on the existence of a one-parameter family of metrics that approaches the effective background metric in a certain way. Although it is not necessary to know this family in an exact form to predict proper\-ties of the backreaction effect, it would be instructive to find explicit examples. In this paper, we provide the first example of such a family of exact nonvacuum solutions to the Einstein equations. It belongs to the Wainwright-Marshman class and satisfies all of the assumptions of the Green-Wald framework. 
\end{abstract}
\section{Introduction}

There is an ongoing debate on the effect of small-scale inhomogeneities on the global structure of space-time, i.e.,\ {\it backreaction}. This is especially interesting in light of cosmological observations which indicate an accelerated expansion of the Universe. Many approaches to the problem have been proposed. Some of them are mathematically rigorous but hardly tractable. Others are tractable but not mathematically rigorous. The approach presented by Green and Wald \cite{greenwald} seems to be, in our opinion, a promising line of research. It is based on Burnett's approach \cite{burnett} and extends it to nonvacuum space-times. The Burnett scheme is itself a mathematically rigorous version of the Isaacson work \cite{isaacson1,isaacson2}.

The Green-Wald framework relies on the existence of a one-parameter family of metrics $g_{ab}(\lambda)$. Such a family depends on a single parameter $\lambda$ and converges uniformly to the background metric $g^{(0)}_{ab}:=\lim_{\lambda\rightarrow 0} g_{ab}(\lambda)$. However, in contrast to the standard perturbation theory, the derivatives of $[g_{ab}(\lambda)-g^{(0)}_{ab}]$ are only required to be bounded and do not necessarily vanish in the limit $\lambda\rightarrow 0$. This may give rise to additional terms that are interpreted as a contribution to the energy-momentum tensor in the Einstein equations for $g^{(0)}_{ab}$. In other words, the background metric $g^{(0)}_{ab}$ does not necessarily solve the Einstein equations with the original energy content. Remaining terms, if any, arise from the averaging of inhomogeneities and may be moved to the right-hand side of the Einstein equations. It has been proved \cite{greenwald} that this additional contribution to the energy-momentum tensor, denoted with $t^{(0)}_{ab}$, is traceless and satisfies the weak energy condition. Hence, the effect of small inhomogeneities reduces to the effect of high-frequency gravitational radiation. This has profound consequences for cosmology. The small-scale inhomogeneities cannot mimic dark energy and cannot be the source of accelerated expansion. 

The Green-Wald result contradicts another popular approach to the problem of backreaction (Buchert \cite{buchert1,buchert2}). The Buchert approach indicates that inhomogeneities may mimic dark energy. Therefore, at least one of these approaches has to be wrong or the application range is different. Objections have been raised \cite{ellis} that the Green-Wald framework is ultralocal and therefore too restrictive to represent well averaging over finite volumes. However, the issue is far from being settled, and more clarification is needed. In particular, the Green-Wald framework has been invented to treat nonvacuum space-times, but no nonvacuum examples of exact solutions that satisfy Green-Wald assumptions have been presented so far. Our paper fills this gap. To date, only two examples of exact solutions compatible with the Green-Wald framework are available in the literature, and these are vacuum space-times; i.e.,\ the Einstein tensor $G_{ab}[g(\lambda)]$ vanishes for $\lambda>0$. The first vacuum family was provided by Burnett in his original paper \cite{burnett} (high-frequency plane waves).\footnote{Note a misprint in \cite{burnett}. The factor $2$ is missing in the exponents in the metric ($1$) there.} The second example of vacuum solutions has been found recently \cite{greenwald2} (the polarized vacuum Gowdy space-times on a torus).\footnote{In Ref.\ \cite{greenwald2}, one more family of solutions is presented, but it violates the weak energy condition as indicated therein.} In this paper, we provide a family of exact nonvacuum solutions to the Einstein equations that satisfies all of the assumptions of the Green-Wald framework. 

\section{Wainwright-Marshman solutions}

The Wainwright-Marshman class \cite{wmc} is a stiff fluid family of inhomogeneous nondiagonal solutions to the Einstein equations. It is usually interpreted as cosmological models with gravitational waves. The metric is given by
\begin{equation}\label{wmmetric}
g=t^{2m}e^n(-dt^2+dz^2)+t^{1/2}\left[dx^2+(t+w^2)dy^2+2wdxdy\right]\;,
\end{equation}
where $t$ is a time coordinate, $x$, $y$, and $z$ are spatial coordinates
\begin{equation}\label{set}
\begin{array}{cc}
t>0\;,\;\; & -\infty<x,y,z<+\infty\;,
\end{array}
\end{equation} 
and $m$ is a parameter. The symbols $n$ and $w$ denote functions of a single null variable $u=t-z$. The Einstein equations reduce to
\begin{equation}\label{wmc}
n'=(w')^2\;,
\end{equation}
and hence one of the functions $n=n(u)$ or $w=w(u)$ may be chosen arbitrarily.
The energy-momentum tensor has a form of a perfect fluid energy-momentum tensor with a stiff equation of state\footnote{We use geometrized units $c=G=1$.}
\begin{equation}\label{sf}
\rho=p=\frac{1}{8\pi}(m+3/16)t^{-2(m+1)}e^{-n}\;.
\end{equation}
The weak energy condition is satisfied for $m\geq-3/16$. The four-velocity of the fluid has a single nonzero component $v^t=t^{-m}e^{-n/2}$.

The equation of state $\rho=p$ implies that the velocity of sound equals the velocity of light; hence, no material in the Universe could be more stiff \cite{icm}. This kind of fluid was probably first proposed by Zeldovich \cite{zel61,zel62} to model the Universe at very high densities just after the big bang. The Wainwright-Marshman solutions are $A_2$ symmetric according to the Wainwright classification scheme \cite{wainclass}. There are two spacelike commuting Killing fields: $\partial_x$ and $\partial_y$. The stiff fluid solutions with such a symmetry may be generated from vacuum solutions with the same symmetry \cite{wimar}.

\section{Inhomogeneity effect}

In order to construct a one-parameter family of metrics satisfying all of the assumptions\footnote{Assumptions {\it (i)-(iv)} of Sec.\ II in Ref.\ \cite{greenwald}.} of the Green-Wald framework, we choose
\begin{equation}\label{w}
w=\lambda \sin \frac{t-z}{\lambda}\;,
\end{equation}
where $\lambda>0$ is a free parameter.
Using the Einstein equation (\ref{wmc}), we have 
\begin{equation}\label{wmn}
n=\frac 1 2 \left(t-z+\frac 1 2 \lambda \sin \frac{2(t-z)}{\lambda}\right)\;,
\end{equation}
where the additive integration constant was set to zero for simplicity. 

The Wainwright-Marshman metric (\ref{wmmetric}) with $w$ given by (\ref{w}), $n$ given by (\ref{wmn}), and $m\geq-3/16$ constitutes a one-parameter family of solutions to the Einstein equations. It is parametrized by a single parameter $\lambda$, and we will denote it with $g_{ab}(\lambda)$. For $m=-3/16$ it corresponds to a vacuum solution.\footnote{The Wainwright-Marshman solutions with $m=-3/16$, $w=const$ belong to the Kasner class \cite{solitons}.} For $m>-3/16$ it corresponds to stiff fluid solutions with the weak energy condition satisfied. 

The background metric $g_{ab}^{(0)}:=\lim_{\lambda\rightarrow 0}g_{ab}(\lambda)$ has a diagonal form:
\begin{equation*}
g^{(0)}=t^{2m}e^\frac{t-z}{2}(-dt^2+dz^2)+t^{1/2}\left(dx^2+tdy^2\right)\;,
\end{equation*}
and it does not belong to the Wainwright-Marshman class, because the functions $w^{(0)}:=\lim_{\lambda\rightarrow 0}w=0$ and $n^{(0)}:=\lim_{\lambda\rightarrow 0}n=\frac{t-z}{2}$ do not satisfy (\ref{wmc}). This implies that one may expect a nonzero backreaction effect here ($t^{(0)}_{ab}\neq 0$).

It is a well-known fact that taking limits is a gauge-dependent procedure \cite{gerochlimit}. Therefore, it is instructive to explain in which sense the effective energy-momentum tensor $t^{(0)}$ is gauge independent \cite{burnett}.\footnote{The standard perturbation theory is gauge independent in a similar way.} As an example, let us consider a coordinate transformation that is valid for $\lambda>0$ but changes the background metric $g^{(0)}$ and the effective energy-momentum tensor $t^{(0)}$. This coordinate transformation alters the physical meaning of the limit $\lambda\rightarrow 0$. In the new coordinates, the one-parameter family of solutions does not model small-scale inhomogeneities anymore (for $\lambda\ll 1$). Let us rewrite $g(\lambda)$ [the Wainwright-Marshman solution (\ref{wmmetric}) with $w$ and $n$ given by (\ref{w}), (\ref{wmn})] by using the following change of coordinates
\begin{eqnarray}\label{ct}
t&=&\lambda\tilde t\;,\\\nonumber
x&=&\lambda^{-1/4}\tilde x\;,\\\nonumber
y&=&\lambda^{-3/4}\tilde y\;,\\\nonumber
z&=&\lambda\tilde z+4(m+1)\ln\lambda\;,
\end{eqnarray}
where $\lambda>0$. The range of the new coordinates $(\tilde t,\tilde x,\tilde y,\tilde z)$ is the same as in (\ref{set}). The metric in the new coordinates is denoted with $\tilde {g}(\lambda)$. The metric $\tilde {g}(\lambda)$ has the Wainwright-Marshman form (\ref{wmmetric}) with $t$, $x$, $y$, $z$, $w$, $n$ substituted by $\tilde t$, $\tilde x$, $\tilde y$, $\tilde z$, $\tilde w$, $\tilde n$, where
\begin{eqnarray*}
\tilde w&=&\sqrt \lambda \sin\left(\tilde t-\tilde z-4(m+1)\frac{\ln \lambda}{\lambda} \right)\;,\\
\tilde n&=&\frac \lambda 2 \left\{\tilde t-\tilde z+\frac 1 2 \sin\left[2\left(\tilde t-\tilde z-4(m+1)\frac{\ln \lambda}{\lambda} \right)\right]\right\}\;.
\end{eqnarray*}
We calculate $\tilde w^{(0)}:=\lim_{\lambda\rightarrow 0}\tilde w=0$, $\tilde n^{(0)}:=\lim_{\lambda\rightarrow 0}\tilde n=0$, and finally $\tilde {g}_{ab}^{(0)}:=\lim_{\lambda\rightarrow 0}\tilde {g}_{ab}(\lambda)$:
\begin{equation*}
\tilde {g}^{(0)}=\tilde{t}^{2m}(-d\tilde{t}^2+d\tilde{z}^2)+\tilde{t}^{1/2}\left(d\tilde{x}^2+\tilde{t}d\tilde{y}^2\right)\;.
\end{equation*}
Therefore, $\tilde {g}_{ab}^{(0)}\neq {g}_{ab}^{(0)}$ and both metrics are not related by a coordinate transformation as it follows from their forms and their Ricci scalars (the Ricci scalars will be analyzed below). Since the metric $\tilde {g}_{ab}^{(0)}$ belongs to the Wainwright-Marshman family, the backreaction effect is absent and $\tilde t^{(0)}_{ab}=0$ [the metric functions $\tilde w^{(0)}$ and $\tilde n^{(0)}$ satisfy (\ref{wmc})]. The Green-Wald formalism is gauge independent provided the coordinate transformations do not change the physical meaning of the limit $\lambda\rightarrow 0$. This seems to be a very natural restriction ---  Burnett's proof \cite{burnett} of the gauge independence of $t^{(0)}_{ab}$ is restricted to the one-parameter family of diffeomorphisms $\phi_\lambda$ that reduce to identity for $\lambda=0$. This condition has been violated by the transformation (\ref{ct}).

In the remaining part of this section, we analyze properties of the metrics $g(\lambda)$ and $g^{(0)}$ and calculate the backreaction effect. 
The determinants and the Ricci scalars of these metrics may be written in the form 
\begin{eqnarray}\nonumber
\det[g(\lambda)]&=&-t^{2(2m+1)}e^{2n}\;,\\\nonumber
R&=&-\frac{1}{8}(16m+3)t^{-2(m+1)}e^{-n}\;,
\end{eqnarray}
where the function $n$ should be substituted by $n^{(0)}$ for $g^{(0)}$.
The determinants are strictly negative in the region of our interest [given by (\ref{set})]. The components of these metrics are obviously nonsingular there; hence, they are smooth metrics of Lorentzian signature on (\ref{set}). The Ricci scalars and the energy density (\ref{sf}) blow up in the limit $t\rightarrow 0$ for all nonvacuum solutions that satisfy the weak energy condition $m>-3/16$ [assuming that $z(t)$ is bounded from below in this limit which is true along any causal curve]. The nature of this initial big bang singularity depends on the value of $m$; see \cite{pulse}. 

For our choice of $w$ function, the Ricci scalar blows up also for $t-z\rightarrow -\infty$. This put some doubts on a cosmological interpretation of these solutions; however, we do not insist on having one. The framework we are interested in should work for any space-time. 
For $t=const$ hypersurfaces $z=+\infty$ are at a finite spatial distance from any finite $z=z_0$, because 
\begin{equation*}
\int_{z_0}^{+\infty} t^{m}e^{n/2} dz
\end{equation*}
is finite. Therefore, one may suspect that the space-time is singular there. However, the limit $t-z\rightarrow -\infty$ cannot be achieved along any causal curve, because $t>0$ (we have the big bang singularity at $t=0$). The Ricci scalar blows up at $z=+\infty$. The space-times with metrics $g(\lambda)$ and $g^{(0)}$  seem to be geodesically future complete with a curious property of curvature blowing up at $z=+\infty$. All causal geodesics are past incomplete and terminate at the curvature singularity $t=0$. The nature of the hypersurfaces $z=\pm\infty$ needs further investigation, but these studies are not in the scope of this paper.

Finally, we calculate $t^{(0)}_{ab}$ and show that it does not vanish.
Following Ref.\ \cite{greenwald}, we introduce $h_{ab}(\lambda):=g_{ab}(\lambda)-g_{ab}^{(0)}$. 
It may be verified by inspection that the Green-Wald framework assumptions (i)-(iv) (see \cite{greenwald}) are satisfied for $g_{ab}(\lambda)$ and $h_{ab}(\lambda)$.

The easiest way to determine the effective energy-momentum tensor $t^{(0)}_{ab}$ is to calculate
\begin{equation*}
t^{(0)}_{ab}=\frac{1}{8\pi}G_{ab}(g^{(0)})-T^{(0)}_{ab}\;,
\end{equation*}
where $T^{(0)}_{ab}=\wlim_{\lambda\rightarrow 0}T_{ab}(\lambda)$.\footnote{The symbol $\wlim$ denotes the {\it weak
limit} in the sense defined in Ref.\ \cite{greenwald}. For our $T_{ab}(\lambda)$, it reduces to the ordinary limit.} The nonzero
components of $T^{(0)}_{ab}$ are
\begin{eqnarray}\nonumber
T^{(0)}_{tt}&=&T^{(0)}_{zz}=\frac{m+\frac{3}{16}}{8\pi}t^{-2},\\\nonumber
T^{(0)}_{xx}&=&\frac{1}{t}T^{(0)}_{yy}=e^{-\frac{t-z}{2}}t^{\left(-2m+\frac 1 2 \right)}T^{(0)}_{tt}\;,
\end{eqnarray}
and, hence, for $t^{(0)}_{ab}$ we have
\begin{equation}\label{t0}
t^{(0)}_{tt}=t^{(0)}_{zz}=-t^{(0)}_{tz}=-t^{(0)}_{zt}=\frac{1}{32\pi t}\;.
\end{equation}
The remaining components of $t^{(0)}_{ab}$ vanish.
Therefore, the effective energy-momentum tensor $t^{(0)}_{ab}$ is traceless and satisfies the weak energy condition, as expected (see theorems in Ref.\ \cite{greenwald}).

One of the assumptions of the Green-Wald formalism is the existence of a smooth tensor field $\mu_{abcdef}$:
\begin{equation}\label{mu}
\mu_{abcdef}=\wlim_{\lambda\rightarrow 0}\left[\nabla_ah_{cd}(\lambda)\nabla_bh_{ef}(\lambda)\right].
\end{equation}
It has the following symmetries \cite{greenwald}:
\begin{equation}\label{musym}
\mu_{abcdef}=\mu_{(ab)(cd)(ef)}=\mu_{abefcd}\;.
\end{equation}
This tensor field encodes the inhomogeneity effect and provides an alternative but more laborious procedure to calculate the effective energy-momentum tensor $t^{(0)}_{ab}$. Following Refs.\ \cite{burnett,greenwald},\footnote{Note a misprint in Ref.\ \cite{greenwald}. The metric $g^{(0)}_{ab}$ is missing there.}
\begin{eqnarray}\label{t0mu}
  8\pi t^{(0)}_{ab} &=&\frac{1}{8}\left(-\mu^{c\phantom{c}de}_{\phantom{c}c\phantom{de}de}-\mu^{c\phantom{c}d\phantom{d}e}_{\phantom{c}c\phantom{d}d\phantom{e}e}+2\mu^{cd\phantom{c}e}_{\phantom{cd}c\phantom{e}de}\right)g^{(0)}_{ab}+\frac{1}{2}\mu^{cd}_{\phantom{cd}acbd}-\frac{1}{2}\mu^{c\phantom{ca}d}_{\phantom{c}ca\phantom{d}bd}\nonumber\\
  &&+\frac{1}{4}\mu^{\phantom{ab}cd}_{ab\phantom{cd}cd}-\frac{1}{2}\mu^{c\phantom{(ab)c}d}_{\phantom{c}(ab)c\phantom{d}d}+\frac{3}{4}\mu^{c\phantom{cab}d}_{\phantom{c}cab\phantom{d}d}-\frac{1}{2}\mu^{cd}_{\phantom{cd}abcd}\,.
\end{eqnarray}
One may find $\mu_{abcdef}$ using (\ref{mu}) by a direct calculation. 
After some algebra we find that the nonzero independent components of $\mu_{abcdef}$ are
\begin{align}\nonumber
&\mu_{tttttt}=\mu_{ttzzzz}=-\mu_{ttttzz}=\mu_{zztttt}=\mu_{zzzzzz}=-\mu_{zzttzz}\\\nonumber
&=-\mu_{tztttt}=-\mu_{tzzzzz}=\mu_{tzttzz}=\frac{1}{8}t^{4m}e^{t-z}\;,\\
&\mu_{ttxyxy}=\mu_{zzxyxy}=-\mu_{tzxyxy}=\frac{t}{2}\;.\label{muco}
\end{align}
The remaining components follow from symmetries (\ref{musym}) or vanish.
It is easy to check that (\ref{muco}) and (\ref{t0mu}) lead to (\ref{t0}). 

One may repeat the calculations in a gauge that leads to a different background space-time $\tilde g^{(0)}$. Similarly, we have $\tilde h_{ab}(\lambda):=\tilde g_{ab}(\lambda)-\tilde g_{ab}^{(0)}$. Again, it may be verified by inspection that the Green-Wald framework assumptions (i)-(iv) (see Ref.\ \cite{greenwald}) are satisfied for $\tilde g_{ab}(\lambda)$ and $\tilde h_{ab}(\lambda)$ (this time they are trivially satisfied). We obtain $\tilde t^{(0)}_{ab}=0$ as predicted before.

Our calculations have been verified with the help of the computer algebra system {\sc Mathematica}.

\section{Summary}

We have presented the first explicit example of a one-parameter family of exact nonvacuum metrics that satisfies all of the Green-Wald assumptions \cite{greenwald}. This example illustrates the backreaction effect and its description within the Green-Wald framework. It provides a convenient setting to study the effect of small-scale inhomogeneities on the large-scale structure of space-time. The effective energy-momentum tensor is traceless and satisfies the weak energy condition, in accord with the theorems of Green and Wald. In particular, the effect of small inhomogeneities on the global structure of space-time cannot mimic a positive cosmological constant or other hypothetical forms of dark energy. This conclusion is not surprising, because there exist other arguments \cite{pulse} to support the hypothesis that inhomogeneities in the model are due only to gravitational waves. Our example provides a starting point for further analysis.

\vspace{0.5cm}

\noindent{\sc Acknowledgments.}
We thank Michał Eckstein and Leszek Sokołowski for discussion and Piotr Chruściel, Alexander Feinstein, Stephen Green, Mikołaj Korzyński, Woei Chet Lim, and Robert Wald for comments. Some calculations were carried out with {\sc Mathematica} and {\sc CCGRG}, {\sc xAct} packages \cite{ccgrg,xAct}. A.K., M.J.W. and S.J.S.\ were supported by the Polish Ministry of Science and Higher Education (the Iuventus Plus Grant No.\ IP2011055071). K.G.\ was supported by the John Templeton Foundation.

\bibliographystyle{unsrt}
\bibliography{report}

\begin{thebibliography}{10}

\bibitem{greenwald}
S.~R. Green and R.~M. Wald.
\newblock New framework for analyzing the effects of small scale
  inhomogeneities in cosmology.
\newblock {\em Phys. Rev. D}, 83:084020, Apr 2011.

\bibitem{burnett}
G.~A. {Burnett}.
\newblock {The high-frequency limit in general relativity}.
\newblock {\em J. Math. Phys.}, 30:90, 1989.

\bibitem{isaacson1}
R.~A. Isaacson.
\newblock {Gravitational Radiation in the Limit of High Frequency. I. The
  Linear Approximation and Geometrical Optics}.
\newblock {\em Phys. Rev.}, 166:1263--1271, Feb 1968.

\bibitem{isaacson2}
R.~A. Isaacson.
\newblock {Gravitational Radiation in the Limit of High Frequency. II.
  Nonlinear Terms and the Effective Stress Tensor}.
\newblock {\em Phys. Rev.}, 166:1272--1280, Feb 1968.

\bibitem{buchert1}
T.~Buchert.
\newblock On average properties of inhomogeneous fluids in general relativity:
  Dust cosmologies.
\newblock {\em General Relativity and Gravitation}, 32:105--125, 2000.
\newblock 10.1023/A:1001800617177.

\bibitem{buchert2}
T.~Buchert.
\newblock On average properties of inhomogeneous fluids in general relativity:
  Perfect fluid cosmologies.
\newblock {\em General Relativity and Gravitation}, 33:1381--1405, 2001.
\newblock 10.1023/A:1012061725841.

\bibitem{ellis}
G.~F.~R. Ellis.
\newblock Inhomogeneity effects in cosmology.
\newblock {\em Classical and Quantum Gravity}, 28(16):164001, 2011.

\bibitem{greenwald2}
S.~R. Green and R.~M. Wald.
\newblock Examples of backreaction of small-scale inhomogeneities in cosmology.
\newblock {\em Phys. Rev. D}, 87:124037, Jun 2013.

\bibitem{wmc}
J.~{Wainwright} and B.~J. {Marshman}.
\newblock {Some exact cosmological models with gravitational waves}.
\newblock {\em Physics Letters A}, 72:275--276, July 1979.

\bibitem{icm}
A.~{Krasi\'nski}.
\newblock {\em {Inhomogeneous Cosmological Models}}.
\newblock Cambridge University Press, July 1997.

\bibitem{zel61}
Ya.~B. Zeldovich.
\newblock {\em ZhETF}, 41:1609, 1961.

\bibitem{zel62}
Ya.~B. Zeldovich.
\newblock The equation of state at ultrahigh densities and its relativistic
  limitations.
\newblock {\em Soviet Physics JETP}, 14:1143, 1962.

\bibitem{wainclass}
J.~Wainwright.
\newblock A classification scheme for non-rotating inhomogeneous cosmologies.
\newblock {\em Journal of Physics A: Mathematical and General}, 12(11):2015,
  1979.

\bibitem{wimar}
J.~{Wainwright}, {Ince}~W. C., and B.~J. {Marshman}.
\newblock {{S}patially {H}omogeneous and {I}nhomogeneous {C}osmologies with
  {E}quation of {S}tate $p=\mu$}.
\newblock {\em Gen.\ Rel.\ Grav.\ }, 10:259--271, 1979.

\bibitem{solitons}
V.~Belinski and E.~Verdaguer.
\newblock {\em {Gravitational Solitons}}.
\newblock Cambridge University Press, 2001.

\bibitem{gerochlimit}
R.~Geroch.
\newblock Limits of {S}pacetimes.
\newblock {\em Commun. math. {P}hys.}, 13:180--193, 1969.

\bibitem{pulse}
J.~Wainwright.
\newblock Gravitational wave pulse in a spatially homogeneous universe.
\newblock {\em Phys. Rev. D}, 20:3031--3038, Dec 1979.

\bibitem{ccgrg}
A.~Woszczyna et~al.
\newblock {ccgrg}: {G}eneral {R}elativity {P}ackage for {M}athematica.
\newblock \url{http://msu.drac.oa.uj.edu.pl/publications.php?pub_content=5}.

\bibitem{xAct}
J.~M. Mart\'{\i}n-Garc\'{\i}a.
\newblock {xAct}: Efficient tensor computer algebra.
\newblock \url{http://www.xact.es}.

\end{thebibliography}

\end{document}